\documentclass[prd,a4paper,preprint,byrevtex]{revtex4}
\usepackage{graphicx}
\usepackage{amssymb}

\begin{document}

\title{Accurate polynomial interpolations of special functions}

\author{Claude \surname{Semay}}
\thanks{FNRS Research Associate}
\email[E-mail: ]{claude.semay@umh.ac.be}
\affiliation{Groupe de Physique Nucl\'{e}aire Th\'{e}orique,
Universit\'{e} de Mons-Hainaut,
Acad\'{e}mie universitaire Wallonie-Bruxelles,
Place du Parc 20, BE-7000 Mons, Belgium}

\date{\today}

\begin{abstract}
Provided a special function of one variable and some of its
derivatives can be accurately computed over a finite range, a method
is presented to build a series of polynomial approximations of the
function with a defined relative error over the whole range. This
method is easy to implement and makes possible fast computation of
special functions.
\end{abstract}


\maketitle

\section{Introduction}
\label{sec:intro}

It is often necessary to compute with a high precision special
functions of one variable within a finite range of values. This task
can be very difficult and can require a great computational time if
the function is known, for instance, by an integral representation or
by a very long expansion. Such functions can be evaluated with a very
high precision by symbolic manipulation languages, but this is not a
very practical method if you need to perform calculations in a Fortran
code for instance.

The idea of the method presented here is to compute the function
considered and some of its derivatives for a special set of points
within the range of interest. This can be performed by any mean:
symbolic manipulation languages or usual computational codes. The
relative accuracy required for the function determines completely the
number of points and their positions within the finite range. Once
this set of points is calculated, the function at any value within the
interval can be computed with the required relative accuracy using
only the information about the function at the point immediately
below and the point immediately above the value. This is possible by
computing a polynomial whose values and values of some of its
derivatives are equal to the corresponding values for the function to
interpolate, for the pair of successive points.

\section{Interpolation with first derivative}
\label{sec:first}

Let us assume that we know exactly a function $F$ and its first
derivative $F'$ at two points $x_1$ and $x_2$. We can easily determine
the third degree polynomial $P(x)$ such that $P(x_1)=F(x_1)$,
$P(x_2)=F(x_2)$, $P'(x_1)=F'(x_1)$, and $P'(x_2)=F'(x_2)$. The
coefficients of the interpolating polynomial can be determined by
solving a Vandermonde-like system \cite{pres92}, but such a system can
be quite ill-conditioned. It is preferable to compute directly $P(x)$
by a Lagrange-like formula \cite{bory73}. Actually, the polynomial
$P(x)$ which satisfies the conditions above is simply given by
\begin{eqnarray}
P(x)&=&  F(x_1)\,f\left( \frac{x-x_1}{x_2-x_1} \right) +
         F(x_2)\,f\left( \frac{x-x_2}{x_1-x_2} \right) \nonumber \\
\label{gpx3}
    &+&  (x_2-x_1)
         \left[ F'(x_1)\,g\left( \frac{x-x_1}{x_2-x_1} \right)
                        -F'(x_2)\,g\left( \frac{x-x_2}{x_1-x_2}
                        \right)
         \right],
\end{eqnarray}
provided the spline polynomials $f$ and $g$ are characterized by the
boundary properties given in Table~\ref{tab:fg}. The expressions
(\ref{fgx}) of these spline functions are given in the Appendix.

\begin{table}[h]
\protect\caption{Boundary properties of the spline functions $f$ and
$g$ for a third degree interpolating polynomial.}
\label{tab:fg}
\begin{tabular*}{\textwidth}{@{\extracolsep{\fill}}ccccc}
\hline
$S(x)$ & $S(0)$ & $S(1)$ & $S'(0)$ & $S'(1)$ \\
\hline
$f(x)$ & 1 & 0 & 0 & 0 \\
$g(x)$ & 0 & 0 & 1 & 0 \\
\hline
\end{tabular*}
\end{table}

\begin{figure}[ht]
\centerline{\includegraphics*[height=8cm]{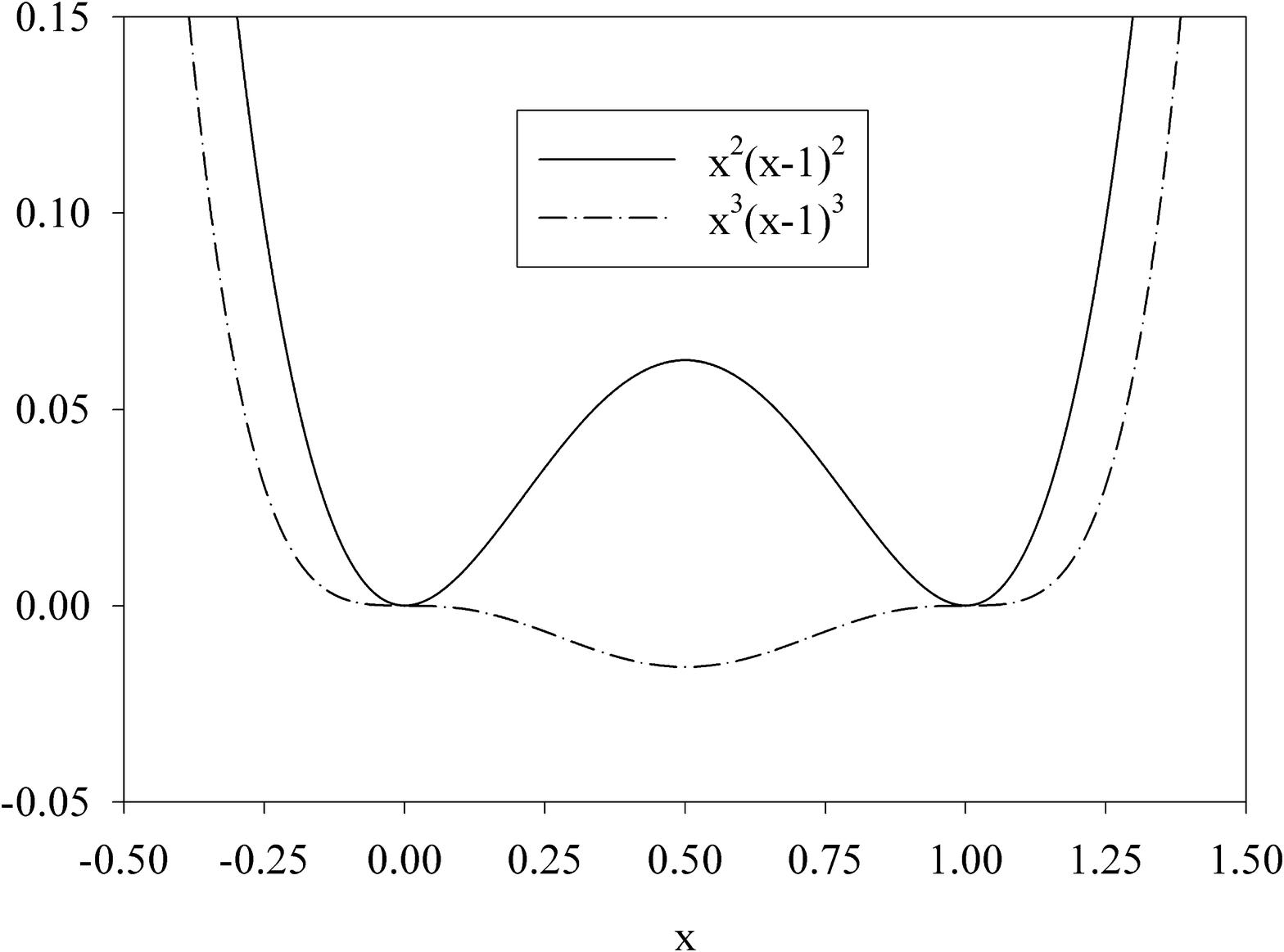}}
\caption{Functions $x^2\left( x-1 \right)^2$ and
$x^3\left( x-1 \right)^3$. In both cases, the extremum appearing
within the interval $[0,1]$ is located at $x=0.5$.}
\label{fig:error}
\end{figure}

It is possible to estimate the error made by using $P(x)$ instead of
$F(x)$ within the interval $[x_1,x_2]$. To simplify calculations, we
can
perform a translation of the coordinate system in order to fix
$x_1=0$ and $F(x_1)=0$, and a rotation to get $F'(x_1)=1$, for
instance. If we note $x_2=h$, $F(x_2)=y$ and $F'(x_2)=z$, the
interpolating polynomial $P(x)$ is given by
\begin{equation}
\label{px}
P(x)=x+\frac{1}{h^2}\left( 3y-zh-2h \right)x^2
+\frac{1}{h^3}\left( -2y+zh+h \right)x^3.
\end{equation}
With the same conventions, the limited Taylor
expansion of the function $F$ around $x_1=0$ is written
\begin{equation}
\label{exp}
F(x)=x+\frac{F''(0)}{2}x^2+\frac{F'''(0)}{6}x^3
+\frac{F^{(4)}(0)}{24}x^4+{\cal O}(x^5).
\end{equation}
Computed in $x=x_2=h$, the expression above and its
first derivative give
\begin{eqnarray}
\label{exp2}
F(h)&=y&\approx
h+\frac{F''(0)}{2}h^2+\frac{F'''(0)}{6}h^3+\frac{F^{(4)}(0)}{24}h^4,
\nonumber \\
F'(h)&=z&\approx 1+F''(0)h+\frac{F'''(0)}{2}h^2
+\frac{F^{(4)}(0)}{6}h^3,
\end{eqnarray}
if we neglect contributions of higher order terms.
We can solve this system to calculate $F''(0)$ and $F'''(0)$ as a
function of $h$, $y$, $z$ and $F^{(4)}(0)$. We can then replace these
two values in Eq.~(\ref{exp}). Using Eq.~(\ref{px}), we finally find
\begin{equation}
\label{fmp1}
F(x)-P(x) \approx \frac{F^{(4)}(0)}{24} x^2 \left( x-h \right)^2.
\end{equation}
The function $x^2 ( x-h )^2$ is represented on Fig.~\ref{fig:error}
for $h=1$. Within the interval $[0,h]$, it presents only one maximum
at
$x=h/2$, and decreases monotonically from this maximum toward zero at
$x=0$ and $x=h$. It is then possible to evaluate the maximum error
within the interval $[0,h]$. Returning to the first notations, we find
\begin{equation}
\label{fmp2}
\max_{[x_1,x_2]}|F(x)-P(x)| \approx \frac{|F^{(4)}(x_1)|}{384}
\left( x_1 - x_2 \right)^4,
\end{equation}
the maximal error being located near the middle of the interval.

For a given set of points for which $F(x)$, $F'(x)$ and $F^{(4)}$ are
known, it is then possible to build an interpolating polynomial for
each interval and to estimate the error within each interval. But it
is
possible to use Eq.~(\ref{fmp2}) in a more clever way. Let us assume
that you need an approximation of a function $F$ within an interval
$[a,b]$ with a fixed relative precision $\epsilon$. If you can compute
$F(x)$, $F'(x)$ and $F^{(4)}(x)$ for arbitrary values $x$ within this
range, you can start from $x_1=a$ to determine a point $x_2$ in such a
way that the relative accuracy of the interpolating polynomial defined
by Eq.~(\ref{gpx3}) is around $\epsilon$ within $[x_1,x_2]$. Then, you
can calculate a point $x_3$ from $x_2$ in a similar way, and so on.
The general relation is
\begin{equation}
\label{hest}
x_{i+1} = x_i + \left|
\frac{384\,\epsilon\, F(x_i)}{F^{(4)}(x_i)}
\right|^{1/4}.
\end{equation}
Finally, a point $x_{N+1} \geq b$ is reached.
With the $N+1$ triplets $(x_i,F(x_i),F'(x_i))$, you can
build, using Eq.~(\ref{gpx3}), a polynomial approximation of $F$ on
$N$ intervals with
$N$ different polynomials $P_i(x)$ of the third degree such that
\begin{equation}
\label{relaterr}
\left| \frac{F(x)-P_i(x)}{F(x)} \right| \lesssim \epsilon \quad
\forall x \in [x_i,x_{i+1}] \quad {\rm and} \quad i=1, 2, \ldots, N.
\end{equation}

If you want to compute $F(x)$ with $x$ within the range $[a,b]$, you
have
to localize first the interval $[x_i,x_{i+1}]$ which contains $x$.
Then the calculation at $x$ of the third degree interpolating
polynomial $P_i(x)$ within this interval will give the evaluation of
$F(x)$ with a relative error of $\epsilon$. These two operations can
be performed very fast \cite{pres92}.

\section{Interpolation with first and second derivatives}
\label{sec:second}

If you can compute higher order derivatives of the function $F$, you
can build better polynomial approximations. The fifth degree
polynomial $P(x)$ such that $P(x_1)=F(x_1)$,
$P(x_2)=F(x_2)$, $P'(x_1)=F'(x_1)$, $P'(x_2)=F'(x_2)$,
$P''(x_1)=F''(x_1)$, and $P''(x_2)=F''(x_2)$ is given by
\begin{eqnarray}
P(x)&=&  F(x_1)\,f\left( \frac{x-x_1}{x_2-x_1} \right) +
         F(x_2)\,f\left( \frac{x-x_2}{x_1-x_2} \right) \nonumber \\
    &+&  (x_2-x_1)
         \left[ F'(x_1)\,g\left( \frac{x-x_1}{x_2-x_1} \right)
               -F'(x_2)\,g\left( \frac{x-x_2}{x_1-x_2} \right)
         \right] \nonumber \\
\label{px5}
    &+&  (x_2-x_1)^2
         \left[ F''(x_1)\,k\left( \frac{x-x_1}{x_2-x_1} \right)
               +F''(x_2)\,k\left( \frac{x-x_2}{x_1-x_2} \right)
         \right],
\end{eqnarray}
provided the spline polynomials $f$, $g$ and $k$ are characterized by
the boundary properties given in Table~\ref{tab:fgk}. The expressions
(\ref{fgkx}) of these spline functions are given in the Appendix.

\begin{table}[h]
\protect\caption{Boundary properties of the spline functions $f$, $g$
and $k$ for a fifth degree interpolating polynomial.}
\label{tab:fgk}
\begin{tabular*}{\textwidth}{@{\extracolsep{\fill}}ccccccc}
\hline
$S(x)$ & $S(0)$ & $S(1)$ & $S'(0)$ & $S'(1)$ & $S''(0)$ & $S''(1)$ \\
\hline
$f(x)$ & 1 & 0 & 0 & 0 & 0 & 0 \\
$g(x)$ & 0 & 0 & 1 & 0 & 0 & 0 \\
$k(x)$ & 0 & 0 & 0 & 0 & 1 & 0 \\
\hline
\end{tabular*}
\end{table}

Using the same procedure as in the previous section, the error between
the function and the interpolating polynomial (\ref{px5}) within the
interval $[0,h]$ is estimated at
\begin{equation}
\label{fmp3}
F(x)-P(x) \approx \frac{F^{(6)}(0)}{720} x^3 \left( x-h \right)^3.
\end{equation}
The function $x^3 ( x-h )^3$ is represented on Fig.~\ref{fig:error}
for $h=1$. Within the interval $[0,h]$, it also presents only one
extremum
at $x=h/2$, and tends monotonically from this extremum toward zero at
$x=0$ and $x=h$. With the most general notations, we find
\begin{equation}
\label{fmp4}
\max_{[x_1,x_2]}|F(x)-P(x)| \approx \frac{|F^{(6)}(x_1)|}{46080}
\left( x_1 - x_2 \right)^6,
\end{equation}
the maximal error being located near the middle of the interval. If
you need an approximation of a function $F$ with a relative precision
$\epsilon$ over a fixed range, and if you can compute
$F(x)$, $F'(x)$, $F''(x)$ and $F^{(6)}(x)$ for arbitrary values $x$
within this range, you can define a series of points with the
following relation
\begin{equation}
\label{hest2}
x_{i+1} = x_i + \left|
\frac{46080\,\epsilon\, F(x_i)}{F^{(6)}(x_i)}
\right|^{1/6},
\end{equation}
in such a way that the fifth degree polynomials built with
Eq.~(\ref{px5}) for each interval are an approximation of $F$ with the
relative accuracy $\epsilon$.

It is possible to define better and better polynomial approximations
by using higher order derivatives of the function under study. But
very good results can already be obtained with the use of the first
and second derivatives only.

\section{Application and concluding remarks}
\label{sec:rem}

These techniques are used here to compute an approximation of the
modified Bessel function of integer order $K_0(x)$ \cite{pres92}. For
a fixed range, the number of points decreases if the second derivative
is used to compute the approximation, in supplement of the first
derivative only. It is also possible to reduce the number of points by
smoothing the function to compute. For instance, we have
\begin{equation}
\label{k0}
K_0(x) \approx \sqrt{\frac{\pi}{2}}\, \frac{\exp(-x)}{\sqrt{x}},
\end{equation}
for large values of $x$. If we remove the rapidly varying exponential
part of $K_0(x)$ by computing $\exp(x)\,K_0(x)$, we can reduce
strongly the number of intervals. The gain is even better by computing
an approximation of $\sqrt{x}\,\exp(x)\,K_0(x)$. These results are
illustrated in Table~\ref{tab:bessel}.

\begin{table}[h]
\protect\caption{Number of points necessary to reach a relative
precision of $10^{-10}$ for the function $F(x)$ with $x$ within the
interval $[2,6]$ ($K_0(x)$ is a modified Bessel function).}
\label{tab:bessel}
\begin{tabular*}{\textwidth}{@{\extracolsep{\fill}}ccc}
\hline
$F(x)$ & With first derivative & With first and \\
  &   & second derivatives \\
\hline
$K_0(x)$ & 342 & 41 \\
$\exp(x)\,K_0(x)$ & 121 & 21 \\
$\sqrt{x}\,\exp(x)\,K_0(x)$ & 68 & 15 \\
\hline
\end{tabular*}
\end{table}

In order to remove divergent or rapidly varying behaviors, it is
sometimes interesting to multiply the function $F$ to approximate by a
function $G$ known with a very weak relative error. An approximation
of $F(x)\, G(x)$ is then computed. The relative precision of the
approximation of $F$ is not spoiled by dividing the interpolating
polynomial by the function $G$, since the relative error on a quotient
is the sum of the relative errors of the factors. So, if the relative
precision for
$G(x)$ is very good, the relative error on $F(x)$ is controlled by the
relative error on $F(x)\, G(x)$.

The number of points necessary to reach a fixed precision obviously
increases with the required accuracy. It depends also strongly on the
range of values. This is shown in Table~\ref{tab:bessel2}.

\begin{table}[h]
\protect\caption{Number of points necessary to reach a relative
precision $\epsilon$ with first and second derivatives for the
function $\sqrt{x}\,\exp(x)\,K_0(x)$ within two
intervals ($K_0(x)$ is a modified Bessel function).}
\label{tab:bessel2}
\begin{tabular*}{\textwidth}{@{\extracolsep{\fill}}cccccc}
\hline
$\epsilon$ & $10^{-10}$ & $10^{-11}$ & $10^{-12}$ & $10^{-13}$ &
$10^{-14}$ \\
\hline
$[2,6]$ & 15 & 21 & 30 & 43 & 62 \\
$[6,10]$ & 7 & 10 & 14 & 19 & 28 \\
\hline
\end{tabular*}
\end{table}

The method used here to compute an approximation of a function $F$
over a finite range with a definite precision is useful mainly in two
cases:
\begin{itemize}
\item You need a code to compute the function $F$ in an usual
programming language, but the computation  with a high accuracy of the
function and some of its derivatives is only possible in a symbolic
manipulation language.
\item You can compute the function $F$ and some of its derivatives in
an usual programming language, but the calculation time is
prohibitive. This can be the case if $F$ is known by an integral
representation or by a very long expansion, for instance.
\end{itemize}
In both cases, it is interesting to compute and store the numbers
$x_i$, $F(x_i)$, $F'(x_i)$, etc. to build a polynomial approximation
of $F$. A demo program is available via anonymous FTP on:
{\tt ftp://ftp.umh.ac.be/pub/ftp\_pnt/interp/}.

\section*{Acknowledgments}
The author thanks the FNRS Belgium for financial support.

\appendix

\section{Spline functions}
\label{sec:app}

We give here the spline functions to define the two kinds of
interpolating
polynomials considered in this paper. A third degree interpolating
polynomial is defined with the two polynomial spline functions
\begin{eqnarray}
f(x)&=&2x^3-3x^2+1,  \nonumber \\
\label{fgx}
g(x)&=&x^3-2x^2+x.
\end{eqnarray}
Their boundary properties are given in Table~\ref{tab:fg}.
A fifth degree interpolating polynomial is defined with the three
polynomial spline functions
\begin{eqnarray}
f(x)&=&-6x^5+15x^4-10x^3+1,  \nonumber \\
g(x)&=&-3x^5+8x^4-6x^3+x, \nonumber \\
\label{fgkx}
k(x)&=&\frac{1}{2}\left( -x^5+3x^4-3x^3+x^2 \right).
\end{eqnarray}
Their boundary properties are given in Table~\ref{tab:fgk}.

\end{document}